\documentclass[a4paper,11pt]{article}
\usepackage{pos}

\title{Supernovae Ia, high-redshift probes, and the Hubble tension: current status and future perspectives}

\author*[1,2,3,4]{M. G. Dainotti}
\author[5,6]{B. De Simone}

\affiliation[1]{Division of Science,National Astronomical Observatory of Japan, 2 Chome-21-1 Osawa, Mitaka, Tokyo, 181-8588, Japan}
\affiliation[2]{The Graduate University for Advanced Studies, SOKENDAI, Shonankokusaimura, Hayama, Miura District, Kanagawa, 240-0115, Japan}
\affiliation[3]{Space Science Institutes , 4765 Walnut St Ste B, Boulder, 80301, CO, USA}
\affiliation[4]{Nevada Center for Astrophysics, University of Nevada,89154, 4505 Maryland Parkway, Las Vegas, 80301, NV, USA}
\affiliation[5]{Dipartimento di Fisica "E.R. Caianiello", Università di Salerno, Via Giovanni Paolo II, 132, Fisciano, Salerno, 84084, Italy}
\affiliation[6]{INFN Gruppo Collegato di Salerno- Sezione di Napoli. c/o Dipartimento di Fisica "E.R. Caianiello", Università di Salerno, Via Giovanni Paolo II, 132, Fisciano, Salerno, 84084, Italy}

\emailAdd{maria.dainotti@nao.ac.jp}

\abstract{The Hubble constant ($H_0$) tension is one of the biggest challenges in modern cosmology. This consists of the discrepancy, at around $5\sigma$, between the local value of $H_0$ measured through Supernovae Ia (SNe Ia) constrained with the Cepheids and the value inferred from the observations of Cosmic Microwave Background (CMB) by Planck data. According to the most appealing cosmological models, such as the flat $\Lambda$CDM, the $H_0$ should not vary according to the measurement method or the redshift $z$ of the probe used for estimating it. Thus, many ideas have been proposed in the literature to face this tension. In the current work, we summarize the results obtained with the binned analysis of SNe Ia, showing a decreasing trend for $H_0$ with $z$ with an evolutionary coefficient $\eta \sim 0.01$, and we further discuss the impact of high-$z$ probes such as Gamma-ray Bursts (GRBs) and quasars (QSOs) that allow reaching constraints on the cosmological parameters that will extend the Hubble diagram to high-$z$ values.}

\FullConference{Frontier Research in Astrophysics – IV (FRAPWS2024)\\
 9-14 September 2024\\
Mondello, Palermo, Italy\\}

\begin{document}
\maketitle

\section{Introduction}
Cosmology finds its best interpretation of nature through the so-called flat $\Lambda$CDM, a theoretical framework based on the co-existence of Cold Dark Matter (CDM, non-relativistic), the Dark Energy phase, responsible for the expansion of the universe and parametrized with the constant $\Lambda$, and the absence of geometrical curvature. This model has been considered the best option to describe the structure and the evolution of our universe, given its relative simplicity and its capability of predicting the acceleration expansion phase as proven in \cite{27,22}. Nevertheless, these considerations do not exempt the $\Lambda$CDM model from having major open issues. Among the most relevant, the \emph{Hubble constant tension} (or $H_0$ tension) has sparked much discussion in the scientific community. The $H_0$ describes the current expansion rate of the universe, and the $H_0$ tension arises from the wide discrepancies in its measurements: these discrepancies range from $4$ to $6\,\sigma$ between the late-time measurement of $H_0$ through SNe Ia (calibrated with Cepheids \cite{28} and other local probes) and the early-time measurement of $H_0$ obtained with the Planck CMB measurements \cite{24}. To tackle the Hubble constant tension, several researchers have proposed theories to resolve the tension including alternative Dark Energy models or modified gravity scenarios.
It is important to stress that the reason for such an open problem could lie behind the presence of hidden biases or redshift-evolution trends in the observed probes. 

In this sense, the study of SNe Ia is crucial. These stellar explosions are considered among the best standard candles because of their almost uniform intrinsic luminosity. Furthermore, their presence in the cosmic distance ladder right after the local observations highlights how crucial it is to understand their properties. For these reasons, SNe Ia require dedicated studies on their lightcurve parameters. To this end, the binning approach has proven to be of invaluable support to the identification of possible evolutionary effects of SNe Ia parameters as a function of $z$. This approach has been successfully applied in \cite{30,31,32}, showing how a decreasing trend for $H_0$ is present with the increase of the redshift.

SNe Ia have been recently observed up to $z=2.9$ \cite{23}. Considering that the Last Scattering Surface (from which the CMB is emitted) is at $z=1100$, it is natural to question ourselves about the presence of intermediate cosmological probes at redshift $3<z<1100$ which can expand further the Hubble diagram and give precious hints about cosmology in the dark age of the universe. Two candidate standard candles have shown to be promising in the forthcoming development of high-$z$: GRBs and QSOs. 

GRBs are panchromatic transients generated from the merging of compact objects or the collapse of very massive stars. GRBs have been observed up to $z=9.4$ \cite{4} and their lightcurves are divided into two main phases: the prompt, the highest in energy (emitting in $\gamma$, X, and sometimes optical) that arises from the internal shocks and the afterglow, which is generated from the interaction of the shockwaves with the circumburst medium and is observed in lower energies (X, optical, and sometimes radio). In some GRB lightcurves, it is possible to observe the plateau phase, namely a flat part of the lightcurve at the beginning of the afterglow which arises from the fallback of materials onto a black hole \cite{49,50,51,52,53} or the spin-down phase of a newborn magnetar \cite{29,26}. The plateau phase is morphologically more regular than the prompt and is the protagonist of interesting astrophysical correlations such as the \emph{Dainotti 3D relation} or \emph{fundamental plane relation} \cite{5,6,7,10,12,14,15,16,30}: this relation links the GRBs peak prompt luminosity $L_{peak}$, the end-of-plateau GRB luminosity and rest-frame time, $L_{a}$ and $T^{*}_{a}$, respectively. To standardize a probe that does not show a fixed intrinsic luminosity, the astrophysical correlations are crucial given that they link the luminosity of the probe with other parameters that do not depend on the luminosity itself and can be measured independently. This correlation has been applied with success into cosmological analysis \cite{1,2,3,9,11,17} to estimate the parameters such as the total matter density of the universe $\Omega_{M}$ or the equation of state parameter $w$.

QSOs are highly energetic galactic nuclei whose luminosity is generated by the friction of gas falling onto their central supermassive black holes. QSOs have been observed up to $z=10.1$ \cite{33} and they obey the Risaliti-Lusso relation \cite{34}: this correlation links the X-ray and the UV luminosity of QSOs and can be used as a cosmological tool \cite{20,1}. 
This relation has been corrected for selection biases and redshift evolution \citep{43} and has been applied GRBs and QSOs represent together a future perspective on the high-$z$ cosmology and, with the future observations of these transients, it will be possible to increase the precision on the already known correlations or to discover new tighter correlations, if any.  

This work is divided as follows. In Sections \ref{sec:SNe1} and \ref{sec:SNe2}, we summarize the findings of the binned analysis applied to SNe Ia. In Section \ref{sec:highz}, we describe the recent contributions in the field of high-$z$ cosmology thanks to the support of GRBs and QSOs. In Section \ref{sec:conclusions} the conclusions are reported.

\section{The Hubble tension in the Pantheon sample: part 1}\label{sec:SNe1}
\cite{30} investigate the Hubble constant tension in the Pantheon sample of SNe Ia. The Pantheon is a collection of $1048$ spectroscopically confirmed SNe Ia with a redshift range of $0.01<z<2.26$ \cite{35} which is divided into 3, 4, 20, and 40 equipopulated bins ordered in redshift. A recalibration of the SNe Ia absolute magnitude $M$ is performed to ensure the value of $H_0=73.5\,km\,s^{-1}\,Mpc^{-1}$ in the local SNe (namely, the first bin out of 3, 4, 20, and 40). For each bin, a Markov Chain Monte-Carlo (MCMC) analysis is conducted to estimate the value of $H_0$, leaving it as the only free-to-vary parameter in the posterior distributions. To do so, the likelihood is assumed as the $\chi^2$ test in each bin:

\begin{equation}
    \chi^2=\Delta \mu^{T} \mathcal{C}^{-1} \Delta \mu,
    \label{eq:chi2}
\end{equation}

where the vector $\Delta \mu = \mu_{th}-\mu_{obs}$, $\mu_{th}$ being the distance modulus inferred from the cosmological models, $\mu_{obs}$ the observed one, and $\mathcal{C}$ is the covariance matrix that contains both statistical and systematic uncertainties. We here remind that the $\mu_{th}$ is computed within two main cosmological models: the $\Lambda$CDM and the $w_{0}w_{a}$CDM \cite{36,37}, where the latter assumes a redshift-evolving behavior for the equation of state parameter $w=w(z)$. After the values of $H_0$ are found, a fitting with a decreasing function of the redshift is conducted. The functional form considered for the fit is the following:

\begin{equation}
    H_0(z)=\frac{H'_0}{(1+z)^\eta},
    \label{eq:fitting}
\end{equation}

where $H'_0$ is a free parameter that describes the value of $H_0(z)$ at $z=0$ and $\eta$ is the evolutionary coefficient. According to the values of $\eta$, any decreasing trend of $H_0$ can be highlighted, if present.

\begin{figure}
    \centering
    \includegraphics[width=1\linewidth]{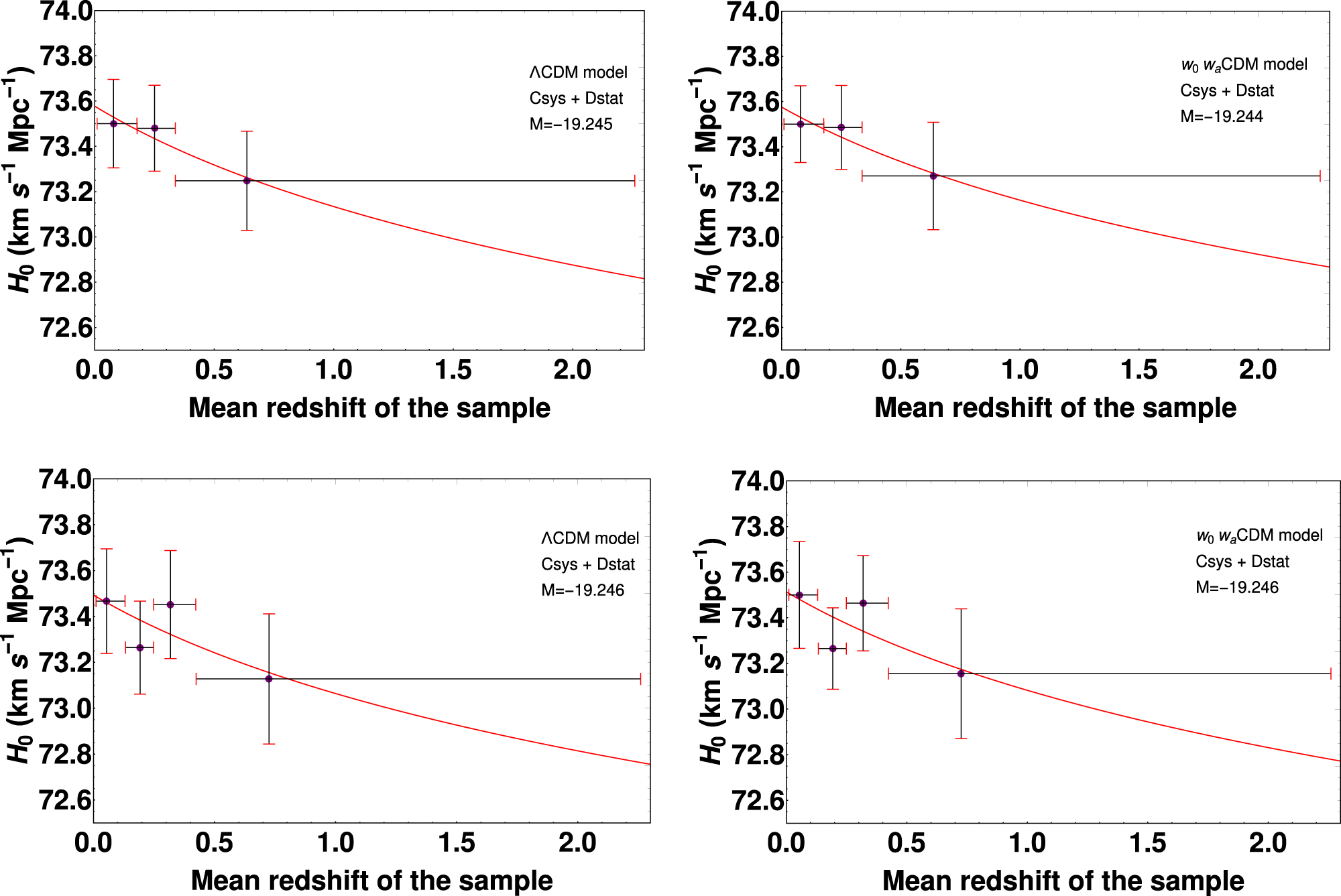}
    \caption{The $H_0(z)$ fitting in the case of 3 and 4 bins extracted from the Pantheon sample. \textbf{Upper left panel.} 3 bins within $\Lambda$CDM. \textbf{Upper right panel.} 3 bins within $w_{0}w_{a}$CDM. \textbf{Lower left panel.} 4 bins within $\Lambda$CDM. \textbf{Lower right panel.} 4 bins within $w_{0}w_{a}$CDM. This Figure is reproduced from \cite{30}.}
    \label{fig:SNe1}
\end{figure}

The results in \cite{30} show that a mild evolution of $H_0$ with $z$ is present. The $\eta$ is $\sim 0.01$ and is compatible with zero in the range from $1.2\sigma$ to $2.0\sigma$. In Figure \ref{fig:SNe1}, we report an example of the fitting procedure for the 3 and 4 redshift-ordered and equipopulated bins of SNe Ia, considering the $\Lambda$CDM and $w_{0}w_{a}$CDM models \cite{30}.

This decreasing trend for $H_0$ with redshift can be explained in light of possible astrophysical biases or redshift-drift for SNe Ia parameters, like what is observed with the stretch, see \cite{38}. In case data are effectively free of any selection effect, alternative cosmological scenarios are suitable to describe the observed results, such as the $f(R)$ modified gravity theories.

\section{The Hubble tension in the Pantheon sample: part 2}\label{sec:SNe2}
\begin{figure}
    \centering
    \includegraphics[width=1\linewidth]{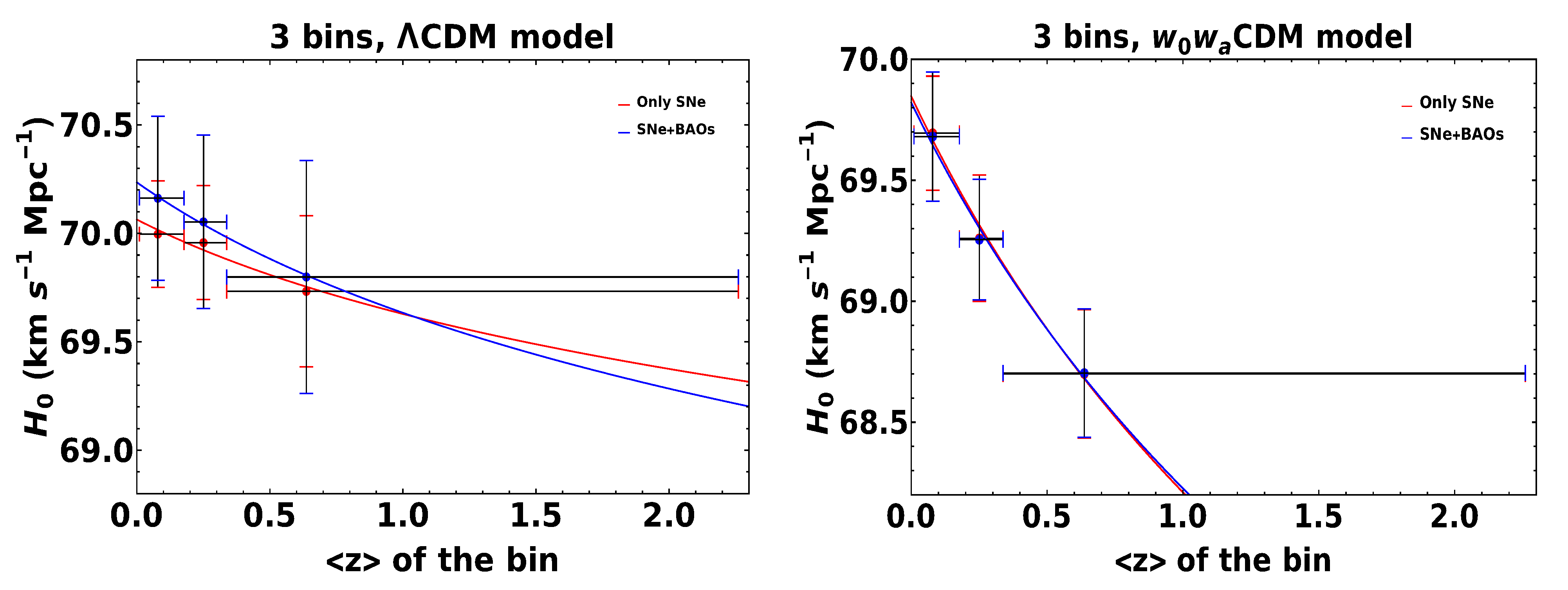}
    \caption{The 3 bins fitting of $H_0(z)$ in the case of Pantheon with two free parameters. \textbf{Left panel.} 3 bins within $\Lambda$CDM. \textbf{Right panel.} 3 bins within $w_{0}w_{a}$CDM. This Figure is reproduced from \cite{31}.}
    \label{fig:SNe2}
\end{figure}

In the second part of the SNe Ia analysis, the purpose is to verify if such a trend could be flattened out through the expansion of the parameters space in the MCMC analysis \cite{31}. The free-to-vary parameters in the $\Lambda$CDM model are $H_0$ and $\Omega_{M}$, while in the $w_{0}w_{a}$CDM model they are $H_0$ and $w_{a}$, this latter being the redshift slope in the CPL parametrization $w(z)=w_{0}+w_{a}\cdot z/(1+z)$. Furthermore, the Baryon Acoustic Oscillations (BAOs, \cite{39}) are added in the SNe Ia likelihood to increase the number of constraints in the estimation of $H_0$. To avoid the statistical fluctuations from dominating the results, the two following preliminary options are adopted: (1) the number of bins is reduced to 3 to have enough SNe Ia in each bin; (2) the re-calibration of SNe Ia is not performed, choosing the default $M=-19.35$ value for Pantheon that corresponds to $H_0=70\,km\,s^{-1}\,Mpc^{-1}$ in the first bin out of 3. After performing the same analysis discussed in Section \ref{sec:SNe1}, the results show $\eta \sim 0.01$, and its compatibility with zero is reached up to $5.8\sigma$. The 3 bins analysis for the Pantheon in two dimensions is shown in Figure \ref{fig:SNe2}, considering both the $\Lambda$CDM and $w_{0}w_{a}$CDM models and the inclusion or exclusion of the BAOs contribution.

A similar decreasing trend for the $H_0$ is observed through the $S$CDM cosmology \cite{40,41}, namely a solution generated by the scale invariance of empty space. \cite{32} prove how the $\eta$ parameter is $\sim 0.01$ for both the Pantheon and Pantheon+ \cite{42} samples, with a compatibility with zero that ranges from 
$5.3\sigma$ to $13.6\sigma$ in the Pantheon and from $4.8\sigma$ to $9.7\sigma$ in the Pantheon+ when the $S$CDM is tested. Such a result could put more constraints on the choice of the adopted cosmological models to tackle the $H_0$ tension or even prove the persistence of hidden astrophysical evolution with $z$ of SNe Ia parameters that are still to be discovered. 

\section{High-redshift probes in aid for cosmology}\label{sec:highz}
In this Section, we summarize some of the most relevant results about cosmology with high-$z$ probes alone and in combination with SNe Ia.

\begin{itemize}
    \item QSOs alone: In \cite{43}, the authors present a selected sample of 983 QSOs, called \emph{golden sample}, extending to redshift $z=7.54$, achieving an intrinsic dispersion of $\delta=0.007$. This precision allows for the determination of the matter density parameter, $\Omega_{M}$, with accuracy comparable to that obtained from SNe Ia (having uncertainties in the order of $0.01$). The selected sample of QSOs can serve as reliable standard candles.  
    \item GRBs alone: \cite{9} perform the simulation of GRBs in the optical and X-ray emission wavelengths to forecast the precision on the cosmological parameter $\Omega_{M}$ that can be achieved with GRBs that follow the Dainotti 3D relation. Considering 134 optical GRBs with the application of machine learning techniques aimed to halve the errors on GRB lightcurve parameters, by 2026 GRBs will be able to reach the same precision on $\Omega_{M}$ as the one of \cite{44}, where $\Delta \Omega_{M} \sim 0.01$. 
    \item QSO and GRBs combined with SNe Ia and BAOs: \cite{1} propose an analysis to estimate the $H_0$ and $\Omega_{M}$ by incorporating GRBs and QSO with SNe Ia and BAOs data to constrain the cosmological parameters. The Gaussianity assumption for the residuals of the theoretical model is tested for all the probes, showing that only the GRBs have a normal distribution in their residuals. Thus, in the other probes, the likelihood is changed according to the true nature of the residuals. This approach leads to precise estimates of the matter density parameter $\Omega_{M}$ and $H_0$, having the uncertainties $\Delta \Omega_{M} \sim 0.001-0.01$ and $\Delta H_0 \sim 0.1$. The improvement in the precision of cosmological parameters allows using these high-$z$ probes in combination with SNe Ia, thus extending the Hubble diagram up to redshifts that are not covered by SNe Ia alone.
\end{itemize}

\section{Conclusions and future perspectives}\label{sec:conclusions}
Modern cosmology is affected by open problems such as the Hubble tension that require the adoption of reliable probes for the estimation of $H_0$. The first step of the cosmological distances ladder is covered by SNe Ia. Given the importance of these explosive events in cosmological analysis, it is fundamental to study the evolution of the $H_0$ with $z$ inside the available SNe Ia samples. Through a binned analysis of SNe Ia, it is possible to observe a mild evolutionary trend for $H_0$ in the Pantheon and Pantheon+ samples, as reported in \cite{30,31,32}. The test of further SNe Ia samples in undergoing \cite{45} and, to confirm any evolutionary trend in SNe Ia, it is crucial to leverage the forthcoming observations of high-$z$ SNe Ia by the Subaru Telescope \cite{46}, the Roman Telescope \cite{47}, the James Webb Space Telescope \cite{48}, and many other facilities both on Earth and in the space.
The relatively limited redshift range for SNe Ia ($z<3$) pushes the scientific community to identify new standardizable candles at redshift ranges that consider $z>3$: in this sense, GRBs and QSOs are among the best candidates given that they are easily observed due to their brightness and due to the existence of reliable astrophysical correlations among their luminosity and other astrophysical parameters. Their combination with the low-$z$ probes has proven to return tight constraints in the estimation of $\Omega_{M}$ and $H_0$: this result stresses the importance of these probes in the immediate future development of modern cosmology.



\begin{thebibliography}{99}
\bibitem{1} G. Bargiacchi et al. "Gamma-ray bursts, quasars, baryonic acoustic oscillations, and supernovae Ia: new statistical insights and cosmological
constraints". In: Monthly Notices of the Royal Astronomical Society 521.3
(May 2023), pp. 3909–3924. doi: 10.1093/mnras/stad763. arXiv: 2303.
07076 {astro-ph.CO}

\bibitem{2} G. Bargiacchi, M. G. Dainotti, and S. Capozziello.
"High-redshift cosmology by Gamma-Ray Bursts: An overview". In: New
Astronomy Reviews 100, 101712 (June 2025), p. 101712. doi: 10.1016/
j.newar.2024.101712. arXiv: 2408.10707 {astro-ph.CO}

\bibitem{3} S. Cao, M. G. Dainotti, and B. Ratra. "Gamma-ray burst data strongly favour the three-parameter fundamental plane (Dainotti) correlation over the two-parameter one". In: Monthly Notices of the Royal Astronomical Society 516.1 (Oct. 2022), pp. 1386–1405. doi: 10.1093/mnras/stac2170. arXiv: 2204.08710 {astro-ph.CO}

\bibitem{4} A. Cucchiara et al. "A Photometric Redshift of $z=9.4$ for GRB 090429B".
In: The Astrophysical Journal 736.1, 7 (July 2011), p. 7. doi: 10.1088/0004-637X/736/1/7. arXiv: 1105.4915 {astro-ph.CO}

\bibitem{5} M. G. Dainotti, V. F. Cardone, and S. Capozziello. "A time-luminosity
correlation for $\gamma$-ray bursts in the X-rays". In: Monthly Notices of the
Royal Astronomical Society 391.1 (Nov. 2008), pp. L79–L83. doi: 10 .
1111/j.1745-3933.2008.00560.x. arXiv: 0809.1389 {astro-ph}

\bibitem{6} M. G. Dainotti et al. "A Fundamental Plane for Long Gamma-Ray Bursts with X-Ray Plateaus". In: The Astrophysical Journal Letters 825.2, L20
(July 2016), p. L20. doi: 10.3847/2041-8205/825/2/L20. arXiv: 1604.06840 {astro-ph.HE}

\bibitem{7} M. G. Dainotti et al. "A Study of the Gamma-Ray Burst Fundamental Plane". In: The Astrophysical Journal 848, 2, 88 (Oct. 2017), p. 88. doi:
10.3847/1538-4357/aa8a6b. arXiv: 1704.04908 {astro-ph.HE}


\bibitem{9} M. G. Dainotti et al. "Optical and X-ray GRB Fundamental Planes as cosmological distance indicators". In: Monthly Notices of the Royal Astronomical Society 514.2 (Aug. 2022), pp. 1828–1856. doi: 10.1093/mnras/stac1141. arXiv: 2203.15538 {astro-ph.CO}

\bibitem{10} M. G. Dainotti et al. "Selection Effects in Gamma-Ray Burst Correlations: Consequences on the Ratio between Gamma-Ray Burst and Star Formation Rates". In: The Astrophysical Journal 800.1, 31 (Feb. 2015), p. 31. doi: 10.1088/0004-637X/800/1/31. arXiv: 1412.3969 {astro-ph.HE}

\bibitem{11} M. G. Dainotti et al. "The gamma-ray bursts fundamental plane correlation as a cosmological tool". In: Monthly Notices of the Royal Astronomical Society 518.2 (Jan. 2023), pp. 2201–2240. doi: 10.1093/mnras/stac2752. arXiv: 2209.08675 {astro-ph.HE}

\bibitem{12} M. G. Dainotti et al. "The Optical Two- and Three-dimensional Funda-
mental Plane Correlations for Nearly 180 Gamma-Ray Burst Afterglows
with Swift/UVOT, RATIR, and the Subaru Telescope". In: The Astro-
physical Journals 261.2, 25 (Aug. 2022), p. 25. doi: 10 . 3847 / 1538 -
4365/ac7c64. arXiv: 2203.12908 {astro-ph.HE}


\bibitem{14} M. G. Dainotti et al. "Determination of the Intrinsic Luminosity
Time Correlation in the X-Ray Afterglows of Gamma-Ray Bursts". In:
The Astrophysical Journal 774.2, 157 (Sept. 2013), p. 157. doi: 10.1088/
0004-637X/774/2/157. arXiv: 1307.7297 {astro-ph.HE}

\bibitem{15} M. G. Dainotti et al. "Discovery of a Tight Correlation for
Gamma-ray Burst Afterglows with "Canonical" Light Curves". In: The
Astrophysical Journal Letters 722.2 (Oct. 2010), pp. L215–L219. doi: 10.
1088/2041-8205/722/2/L215. arXiv: 1009.1663 {astro-ph.HE}

\bibitem{16} M. G. Dainotti et al. "Study of Possible Systematics in the L*X-T*a Correlation of Gamma-ray Bursts". In: The Astrophysical Journal
730.2, 135 (Apr. 2011), p. 135. doi: 10.1088/0004- 637X/730/2/135.
arXiv: 1101.1676 {astro-ph.HE}

\bibitem{17} A. Favale et al. "Towards a new model-independent calibration of
Gamma-Ray Bursts". In: Journal of High Energy Astrophysics 44 (Nov.
2024), pp. 323–339. doi: 10.1016/j.jheap.2024.10.010. arXiv: 2402.
13115 {astro-ph.CO}



\bibitem{20} A. L. Lenart et al. "A Bias-free Cosmological Analysis with
Quasars Alleviating H 0 Tension". In: The Astrophysical Journal Supple-
ment 264.2, 46 (Feb. 2023), p. 46. doi: 10 . 3847 / 1538 - 4365 / aca404.
arXiv: 2211.10785 {astro-ph.CO}


\bibitem{22} S. Perlmutter et al. "Measurements of $\Omega$ and $\Lambda$ from 42 High-Redshift
Supernovae". In: The Astrophysical Journal 517.2 (June 1999), pp. 565–
586. doi: 10.1086/307221. arXiv: astro-ph/9812133 {astro-ph}

\bibitem{23} J. D. R. Pierel et al. "Discovery of an Apparent Red, High-velocity Type
Ia Supernova at z = 2.9 with JWST". In: The Astrophysical Journall
971.2, L32 (Aug. 2024), p. L32. doi: 10.3847/2041-8213/ad6908. arXiv:
2406.05089 {astro-ph.GA}

\bibitem{24} Planck Collaboration et al. "Planck 2018 results. VI. Cosmological pa-
rameters". In: Astronomy \& Astrophysics 641, A6 (Sept. 2020), A6. doi:
10.1051/0004-6361/201833910. arXiv: 1807.06209 {astro-ph.CO}


\bibitem{26} N. Rea et al. "Constraining the GRB-Magnetar Model by Means of the
Galactic Pulsar Population". In: The Astrophysical Journal 813.2, 92
(Nov. 2015), p. 92. doi: 10.1088/0004- 637X/813/2/92. arXiv: 1510.
01430 {astro-ph.HE}

\bibitem{27} Adam G. Riess et al. In: "Observational Evidence from Supernovae for an Accelerating Universe and a Cosmological Constant". The Astronomical Journal 116.3 (Sept. 1998),
pp. 1009–1038. doi: 10.1086/300499. arXiv: astro-ph/9805201 {astro-ph}

\bibitem{28} Adam G. Riess et al. "A Comprehensive Measurement of the Local Value of the Hubble Constant with 1 km s-1 Mpc-1 Uncertainty from the Hubble Space Telescope and the SH0ES Team". In: The Astrophysical Journal
934.1, L7 (July 2022), p. L7. doi: 10.3847/2041- 8213/ac5c5b. arXiv:
2112.04510 {astro-ph.CO}

\bibitem{29} A. Rowlinson et al. "Constraining properties of GRB magnetar central
engines using the observed plateau luminosity and duration correlation".
In: Monthly Notices of the Royal Astronomical Society 443.2 (Sept. 2014),
pp. 1779–1787. doi: 10.1093/mnras/stu1277. arXiv: 1407.1053 {astro-ph.HE}


\bibitem{30} M. G. Dainotti et al. "On the Hubble Constant Tension in the SNe Ia Pantheon Sample". In: The Astrophysical Journal 912, 150 (May 2021). doi: 10.3847/1538-4357/abeb73. arXiv: 2103.02117 {astro-ph.CO}

\bibitem{31} M. G. Dainotti et al. "On the evolution of the Hubble constant with the SNe Ia Pantheon Sample and Baryon Acoustic Oscillations: a feasibility study for GRB-cosmology in 2030". In: Galaxies, 10(1), 24 (January 2022). doi: 10.3390/galaxies10010024. arXiv: 2201.09848 {astro-ph.CO}

\bibitem{32} B. De Simone et al. "A doublet of cosmological models to challenge the $H_0$ tension in the Pantheon Supernovae Ia catalog". In: Journal of High Energy Astrophysics, 45, p. 290-298 (Mar. 2025). doi: 10.1016/j.jheap.2024.12.003. arXiv: 2411.05744 {astro-ph.CO}

\bibitem{33} P. Natarajan et al. "First Detection of an Overmassive Black Hole Galaxy UHZ1: Evidence for Heavy Black Hole Seed Formation from Direct Collapse". In: The Astrophysical Journal Letters, 960, L1 (Dec. 2023). doi: 10.3847/2041-8213/ad0e76. arXiv: 2308.02654 {astro-ph.HE}

\bibitem{34} G. Risaliti and E. Lusso "A Hubble Diagram for Quasars". In: The Astrophysical Journal, 815, 33 (Dec. 2015). doi: 10.1088/0004-637X/815/1/33. arXiv: 1505.07118 {astro-ph.CO}

\bibitem{35} D. M. Scolnic et al. 2018 "The Complete Light-curve Sample of Spectroscopically Confirmed SNe Ia from Pan-STARRS1 and Cosmological Constraints from the Combined Pantheon Sample". In: The Astrophysical Journal, 859, 101 (May 2018). doi: 10.3847/1538-4357/aab9bb. arXiv: 1710.00845 {astro-ph.CO}

\bibitem{36} M. Chevallier and D. Polarski, "Accelerating Universes with Scaling Dark Matter." In: International Journal of Modern Physics D, 10, 213–224 (Feb. 2001). doi: 10.1142/S0218271801000822. arXiv: gr-qc/0009008

\bibitem{37} E. V. Linder, "Exploring the Expansion History of the Universe". In: Physical Review Letters, 90, 091301 (2003). doi: 10.1103/PhysRevLett.90.091301. arXiv: 0208512 {astro-ph}

\bibitem{38} N. Nicolas et al. "Redshift evolution of the underlying type Ia supernova stretch distribution". In: Astronomy \& Astrophysics, 649, A74, p.10 (May 2021). doi: 10.1051/0004-6361/202038447. arXiv: 2005.09441 {astro-ph.CO}

\bibitem{39} G. S. Sharov and V. O. Vasiliev, "How predictions of cosmological models depend on Hubble parameter data sets". In: Math. Model. Geom. 6 (2018). doi:10.26456/mmg/2018-611

\bibitem{40} A. Maeder, "An Alternative to the $\Lambda$CDM Model: The Case of Scale Invariance". In: The Astrophysical Journal, 834, 194. doi: 10.3847/1538-4357/834/2/194, arXiv: 1701.03964

\bibitem{41} J. F. Jesus, "Exact Solution for Flat Scale-Invariant Cosmology". In: arXiv e-prints, arXiv:1712.00697 (2017). doi: 10.48550/arXiv.1712.00697

\bibitem{42} D. M. Scolnic et al. "The Pantheon+ Analysis: The Full Dataset and Light-Curve Release". In: The Astrophysical Journal, 938, 113 (Oct. 2022). doi: 10.3847/1538-4357/ac8b7a. arXiv: 2112.03863 {astro-ph.CO}

\bibitem{43} M. G. Dainotti et al. "Quasars: Standard Candles up to $z=7.5$ with the Precision of Supernovae Ia". In: The Astrophysical Journal, 950, 45 (Jun. 2023). doi: 10.3847/1538-4357/accea0. arXiv: 2305.19668 {astro-ph.CO}

\bibitem{44} M. Betoule et al. "Improved cosmological constraints from a joint analysis of the SDSS-II and SNLS supernova samples". In: Astronomy \& Astrophysics, 568, A22, 32 (Aug. 2014). doi: 10.1051/0004-6361/201423413. arXiv: 1401.4064 {astro-ph.CO}
 
\bibitem{45} M. G. Dainotti et al. "A New Master Supernovae Ia sample and the investigation of the $H_0$ tension". Submitted to Journal of High Energy Astrophysics. arXiv: 2501.11772 {astro-ph.CO}

\bibitem{46} K. Kodaira, "Japan National Large Telescope (SUBARU) Project". In: European Southern Observatory Conference and Workshop Proceedings, European Southern Observatory. p. 43 (1998)

\bibitem{47} T. Eifler et al. "Cosmology with the Roman Space Telescope - multiprobe strategies". In: Monthly Notices of the Royal Astronomical Society 507, 1746–1761 (2021). doi: 10.1093/mnras/stab1762. arXiv:2004.05271 

\bibitem{48} J. P. Gardner et al. "The James Webb Space Telescope Mission". In: Publications of the Astronomical Society of the Pacific 135, 068001 (2023). doi:10.1088/1538-3873/acd1b5. arXiv:2304.04869

\bibitem{49} W.-H. Lei et al. "Hyper-accreting black hole as GRB central engine. I: Baryon loading in GRB jets". In: The Astrophysical Journal, 765, 125, 2. doi: 10.1088/0004-637X/765/2/125. arXiv: 1209.4427 {astro-ph.HE}

\bibitem{50} B. Zhang, "The Physics of Gamma-Ray Bursts". In: HEPRO-VII, University of Nevada (Dec. 2018). doi: 10.1017/9781139226530

\bibitem{51} M. G. Dainotti et al. "The Scavenger Hunt for Quasar Samples to Be Used as Cosmological Tools". In: Galaxies, 12(1), 4 (Jan. 2024). doi: 10.3390/galaxies12010004. arXiv: 2401.11998 {astro-ph.CO}

\bibitem{52} M. G. Dainotti, "A new binning method to choose a standard set of Quasars". In: Physics of the Dark Universe, 44, 101428 (Jan. 2024). doi: 10.1016/j.dark.2024.101428. arXiv: 2401.12847 {astro-ph.HE} 

\bibitem{53} B. Zhang et al., "Physical processes shaping gamma-ray burst X-ray afterglow light curves: theoretical implications from the Swift X-ray telescope observations". In: The Astrophysical Journal, 642, 1, 354 (2006). doi: 10.1086/500723. arXiv: astro-ph/0508321



\end{thebibliography}
\end{document}